\documentclass[10pt,twocolumn,letterpaper]{article}

\usepackage{cvpr}
\usepackage{times}
\usepackage{epsfig}
\usepackage{graphicx}
\usepackage{amsmath}
\usepackage{amssymb}

\usepackage[breaklinks=true,bookmarks=false]{hyperref}

\cvprfinalcopy 

\def\cvprPaperID{****} 

\ifcvprfinal\fi
\begin{document}

\newcommand\Mark[1]{\textsuperscript#1}

\title{End-to-End Learning for Video Frame Compression with Self-Attention}
\author{Nannan Zou\Mark{2}\\
{\tt\small nannan.zou.ext@nokia.com}
\and
Honglei Zhang\Mark{1}\\
{\tt\small honglei.1.zhang@nokia.com}\\
\and
Francesco Cricri\Mark{1}\\
{\tt\small francesco.cricri@nokia.com}\\
\and
Hamed R. Tavakoli\Mark{1}\\
{\tt\small hamed.rezazadegan\_tavakoli@nokia.com}\\
\and
Jani Lainema\Mark{1}\\
{\tt\small jani.lainema@nokia.com}\\
\and
Emre Aksu\Mark{1}\\
{\tt\small emre.aksu@nokia.com}\\
\and
Miska Hannuksela\Mark{1}\\
{\tt\small miska.hannuksela@nokia.com}\\
\and
Esa Rahtu\Mark{2}\\
{\tt\small esa.rahtu@tuni.fi}\\
\and
\Mark{1}Nokia Technologies, \Mark{2}Tampere University, Tampere, Finland\\
}

\maketitle

\begin{abstract}
	One of the core components of conventional (i.e., non-learned) video codecs consists of  predicting a frame from a previously-decoded frame, by leveraging temporal correlations. 
   In this paper, we propose an end-to-end learned system for compressing video frames. Instead of relying on pixel-space motion (as with optical flow), our system learns deep embeddings of frames and encodes their difference in latent space. At decoder-side, an attention mechanism is designed to attend to the latent space of frames to decide how different parts of the previous and current frame are combined to form the final predicted current frame. Spatially-varying channel allocation is achieved by using importance masks acting on the feature-channels. 
   The model is trained to reduce the bitrate by minimizing a loss on importance maps and a loss on the probability output by a context model for arithmetic coding. 
   In our experiments, we show that the proposed system achieves high compression rates and high objective visual quality as measured by MS-SSIM and PSNR. Furthermore, we provide ablation studies where we highlight the contribution of different components. 
   
\end{abstract}

\makeatletter
\def\blfootnote{\xdef\@thefnmark{}\@footnotetext}
\makeatother
\blfootnote{\textcopyright 2020 IEEE. Personal use of this material is permitted. Permission from IEEE must be obtained for all other uses, in any current or future media, including reprinting/republishing this material for advertising or promotional purposes, creating new collective works, for resale or redistribution to servers or lists, or reuse of any copyrighted component of this work in other works.}

\section{Introduction}
Traditional video compression methods are mostly based on intra-frame and inter-frame prediction, followed by transform-coding, as in HEVC/H.265 standard \cite{Sullivan2012a}. For inter-frame prediction, motion prediction and motion compensation are performed, where blocks of frames are predicted from blocks of previously-reconstructed reference frames which share similar content, typically nearby frames. In this paper, the predicted frame is referred to as \textit{P-frame} or \textit{current frame} $f_t$, and the reference frame as \textit{previous frame} $f_{t-1}$.
Recently, neural networks have been applied to image and video compression with promising results. These systems typically follow the auto-encoder paradigm, where the encoder and decoder networks operate as non-linear transform and inverse transform, respectively. In this paper, we describe our end-to-end learned P-frame compression system that we submitted to the 2020 Challenge on Learned Image Compression (CLIC), P-frame compression track. Our submission name was \texttt{ntcodec\_r3}. The goal consists of encoding information, using a low bitrate, which allows a decoder to reconstruct $f_t$ given $f_{t-1}$. Instead of relying on pixel-space motion information, as typically done in previous works which consider optical flow, we propose to first extract frame-embeddings followed by encoding differences in latent space. Importance maps are used to assign a spatially-varying number of channels to each spatial location of the features. The probability distribution of symbols to be encoded/decoded via arithmetic coding is modeled by a multi-scale context model, which is learned jointly with all other neural networks. At decoder side, a learned attention mechanism analyzes frame-embeddings to adaptively combine an initial prediction of the current frame with the previous frame.

In \cite{Chen2019a} and \cite{Mentzer2018}, attention mechanisms are proposed for generating importance masks that weigh the features extracted at encoder-side, implicitly adapting the bit allocation for feature elements based on their importance. In \cite{LiM2018}, importance maps needed to be encoded into the bitstream. 

In \cite{choi2019deepframe}, inter-frame prediction of HEVC is improved by using a deep CNN to produce spatially-varying
filters from the decoded frames to synthesize the predicted patch.
In \cite{chen2019learning},
deep learning techniques are applied within an architecture similar
to traditional video codecs. An input image is partitioned
into patches, and a deep CNN with LSTM blocks performs inter-frame and intra-frame prediction. One important aspect in inter-frame prediction is how to model both static and dynamic information. To this end, the Video Ladder Network \cite{Cricri2016} includes \textit{lateral recurrent residual blocks} as part of a recurrent auto-encoder architecture for predicting the next frame given $10$ previous frames.

One common algorithm to achieve lossless compression is arithmetic coding, which requires an estimate of the probability of the next symbol to be encoded/decoded. This estimate is typically provided by a probability model, whose accuracy directly impacts the compression rate. Neural network architectures such as PixelCNN \cite{Salimans2017}
and PixelCNN++ \cite{oord2016pixelrecurrent} may be used for estimating the probability distribution. These are autoregressive models, where masked
convolutions are used to estimate the distribution of a pixel based on already predicted
pixels acting as context. These models are very slow since they run a heavy deep CNN to calculate the parameters
of the probability model for every pixel or sub-pixel to be encoded.
In \cite{balle2018variational}, the parameters of the probability
model are estimated using a shallow CNN to capture the local correlations
and a hyper-prior branch to incorporate the global context. Although
only a shallow CNN is used, this solution is still too inefficient with respect to the decoding time requirements of the CLIC challenge. Recently, a multi-scale probability model was presented
in \cite{mentzer2019practical}. This model estimates the parameters
of the probability distribution for a pixel using a low-resolution representation
of the input image and the same procedure is applied to multiple scales
of the image. This method can achieve fast encoding and
decoding speed with good accuracy. Our probability model is adapted
from this method with some modifications.


\section{Proposed Method}
\label{sec:method}
\subsection{Architecture}
\begin{figure*} 
\hfill{}\includegraphics[width=2.1\columnwidth]{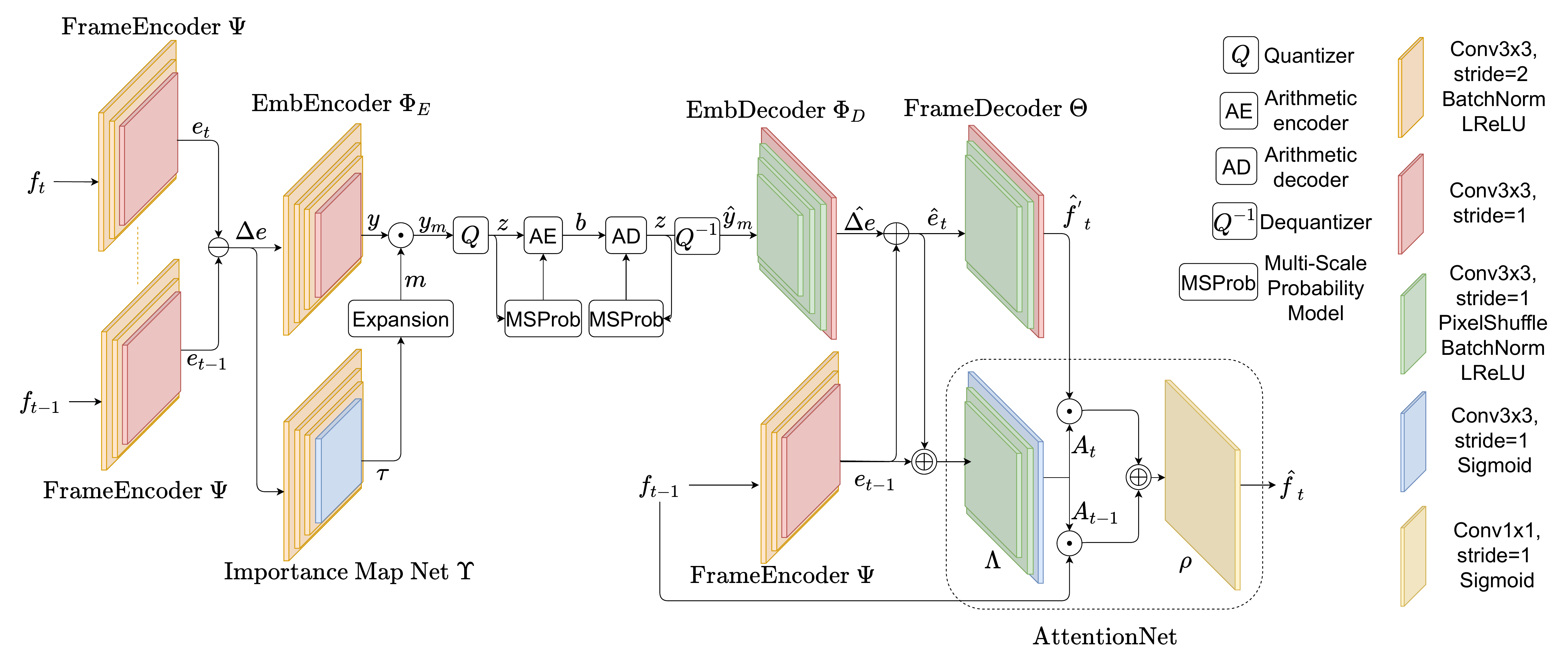}\hfill{} 
\caption{\label{fig:overview}Overview of our proposed solution. PixelShuffle \cite{Shi2016} is used for upsampling at decoder-side.} 
\end{figure*}

An overview of our proposed approach is illustrated in Fig. \ref{fig:overview}.
The basic structure consists of an encoder and a decoder. At encoder side, the two input frames $f_t$ and $f_{t-1}$ are projected into embedded or latent space by two \textit{frame-encoders} neural networks which share weights and are denoted as $\Psi$, obtaining two embeddings $e_t, e_{t-1} \in \mathbb{R}^{\frac{H}{n_\Psi},\frac{W}{n_\Psi},F_{\Psi}}$, where $H,W$ are the height and width of the input frames, $n_\Psi$ is the downsampling factor of $\Psi$, $F_{\Psi}$ is the number of filters in the last layer of $\Psi$. An embedding-difference is then computed:
\begin{equation}
\Delta e = e_t - e_{t-1} = \Psi(f_t)-\Psi(f_{t-1}), 
\label{eq:delta_e}
\end{equation}
which is encoded by an \textit{embedding-encoder} neural network $\Phi_E$. The output latent tensor $y \in \mathbb{R}^{\frac{H}{n_\Psi+n_\Phi}, \frac{W}{n_\Psi+n_\Phi},F_\Phi}$, where $n_\Phi$ is the downsampling factor of $\Phi_E$ and $F_\Phi$ is the number of filters in the last layer of $\Phi_E$, is multiplied by a binary importance mask $m$ (see Section \ref{subsec:importance}) that zeros-out a spatially-varying number of channels:
\begin{equation}
y_m = y \odot m = \Phi_E(\Delta e) \odot m,
\label{eq:masking}
\end{equation}
where $\odot$ indicates element-wise multiplication. The masked output is quantized by 8-bits uniform scalar quantization and then entropy coded by an arithmetic encoder. In order to allow for back-propagating non-zero gradients, we use the straight-through estimator for quantization, as in \cite{Theis2017}. A learned multi-scale context model is used by the arithmetic codec to estimate the probability distribution of next symbols to encode/decode.
At decoder side, the entropy decoded bitstream is dequantized into $\hat{y}$ and input to an \textit{embedding-decoder} $\Phi_D$. The reconstructed embedding for the current frame is obtained as follows:
\begin{equation}
\hat{e}_t = e_{t-1} + \widehat{\Delta e} = \Psi(f_{t-1}) + \Phi_D(\hat{y}).
\label{eq:hat_delta_e}
\end{equation}
However, in a realistic scenario, the decoder would not have the uncompressed version of the previous frame, but only a reconstructed version $\hat{f}_{t-1}$.

An initial version $\hat{f}^{'}_t$ of the reconstructed current frame is obtained by reprojecting $\hat{e}_t$ into pixel space via a \textit{frame-decoder} neural network $\Theta$: 
\begin{equation}
\hat{f}^{'}_t = \Theta(\hat{e}_t)
\label{eq:temp_pred}
\end{equation}
Finally, an attention mechanism is applied to adaptively combine $f_{t-1}$ and $\hat{f}^{'}_t$, as described in Section \ref{subsec:attention}.

\subsection{Learned Spatially-varying Channel Masking}
\label{subsec:importance}
In order to allow the model to allocate a varying number of channels to different spatial areas of the encoded tensor $y$ (see Eq. \eqref{eq:masking}), we use an additional neural network $\Upsilon$ which analyzes the embedding-difference $\Delta e$ and outputs an importance map $\tau \in \mathbb{R}^{\frac{H}{n_\Psi+n_\Phi}, \frac{W}{n_\Psi+n_\Phi},1}$ with elements in $[0,1]$. This map is then quantized with $\log_2 F_\Phi $ bits and then expanded into a mask $m \in \mathbb{R}^{\frac{H}{n_\Psi+n_\Phi}, \frac{W}{n_\Psi+n_\Phi},F_\Phi}$:
\begin{equation}
m_{i,j,k}=
\begin{cases}
1 & \text{if } k < F_\Phi \tau_{i,j} \\
0 & \text{otherwise.}
\end{cases}
\label{eq:expansion}
\end{equation}
In order to encourage masked representations $y_m$ that have low entropy and thus be more easily predictable by our probability model for arithmetic coding, we use the following constraint in our training objective function:
\begin{equation}
\mathcal{M}(\tau) = \big\lvert \bar{\tau}-\beta \big\lvert,
\label{eq:im_loss}
\end{equation}
where $\beta$ is a constant representing the target average non-zero ratio in $m$ and thus in $y_m$.

\subsection{Self-Attention for Adaptive Frame Mixing}
\label{subsec:attention}
We propose a learned self-attention mechanism which allows to adaptively mix information from $f_{t-1}$ and $\hat{f}^{'}_t$. The rationale is that $\hat{e}_t$ may not contain sufficient information for reconstructing all the details in pixel-space, especially when bitrate is constrained. Thus, we relaxed the training by allowing the model to leverage the highly accurate information already present in pixel space of $f_{t-1}$, while using $\hat{f}^{'}_t$ only for the parts which have changed due to motion. This is realized via a self-attention model which is trained jointly with all other neural networks in the system. 
An attention neural network $\Lambda$ analyzes the two embeddings $\hat{e}_t$ and $e_{t-1}$ in order to output an attention map $A_t \in \mathbb{R}^{H,W,3}$, with elements $a_{i,j,k}$ in $[0,1]$: 
\begin{equation}
A_t = \Lambda(\hat{e}_t \oplus e_{t-1}),
\label{eq:attention_map}
\end{equation}
where $\oplus$ indicates tensor concatenation in the channel axis. A second attention map is derived as $A_{t-1}=1-A_t$. The attention maps are applied to the initial reconstructed frame and the previous frame, and the result is then further processed by a 1x1 convolutional layer $\rho$:
\begin{equation}
\hat{f}_t = \rho(A_t \odot \hat{f}^{'}_t + A_{t-1} \odot f_{t-1})
\label{eq:apply_attention}
\end{equation}


\subsection{Probability Model for Arithmetic Encoder}

Our probability model is derived from the multi-scale approach described
in \cite{mentzer2019practical}. 

Let $z^{(i)}$ be the input tensor at scale $i$. We model the distribution
$p\left(z^{(i)}\right)$ conditional to the tensor $z^{(i+1)}$ at
scale $i+1$, i.e., $p\left(z^{(i)}\vert z^{(i+1)}\right)$.
Similar to many previous approaches \cite{balle2018variational,mentzer2019practical,Salimans2017},
we use a generalization of the descretized logistic mixture model
as the distribution model. Let $c$ be the channel index, and $u$,$v$
be the spatial index. We define 
\begin{equation}
p\left(z_{c,u,v}^{(i)}\right)=\sum_{k=1}^{K}\pi_{k,u,v}^{(i)}\sigma\left(\mu_{k,c,u,v}^{(i)},s_{k,c,u,v}^{(i)}\right),\label{eq:logistic_mixture_model}
\end{equation}
where $K$ is the number of mixtures, $\pi_{k,u,v}^{(i)}$ is the
mixture weight parameter, $\sigma(\cdot)$ is the discretized logistic
probability density function, $\mu_{k,c,u,v}^{(i)}$ is the location
parameter, and $s_{k,c,u,v}^{(i)}$ is the scale parameter. Note that
in Eq. \ref{eq:logistic_mixture_model}, the mixture weights $\pi_{k,u,v}^{(i)}$
are shared across all channels of the same spatial location. This
follows the same principle as described in \cite{Salimans2017}.
Taking the channel dependencies into consideration, we let the location
parameter $\mu$ depend on the previous encoded/decoded channel,
such that 
\begin{equation}
\mu_{k,c,u,v}^{(i)}=\tilde{\mu}_{k,c,u,v}^{(i)}+\lambda_{k,c,u,v}^{(i)}z_{c-1,u,v}^{(i)},\label{eq:channel_dependencies}
\end{equation}
where $\tilde{\mu}_{k,c,u,v}^{(i)}$ is location parameter estimated
from $z^{(i+1)}$, $\lambda_{k,c,u,v}^{(i)}$ is a weight parameter
to be learned, and $z_{c-1,u,v}^{(i)}$ is the value from the previous
encoded/decoded channel. Note that in Eq. \ref{eq:channel_dependencies},
we let each channel to be dependent only on the previous channel.
For the first channel, we let $\lambda=0$. This design is critical
when the number of channels is big since it greatly reduces the number
of parameters to be estimated by the network comparing to a fully
dependent mode where each channel depends on all previously decoded
channels. 

For each element $z_{c,u,v}^{(i)}$, the following parameters are
used to describe the distribution, $\tilde{\mu}_{k,c,u,v}^{(i)}$,
$\lambda_{k,c,u,v}^{(i)}$, $s_{k,c,u,v}^{(i)}$, and $\pi_{k,u,v}^{(i)}$.
For a tensor with shape $C\times H\times W$, the total number of
parameters is $C\times3\times K\times H\times W$, where $K$ is
the number of mixtures. These parameters are estimated using a deep
CNN taking $z^{(i+1)}$ as its input. 

\begin{figure}[!htb]
\includegraphics[width=8.5cm]{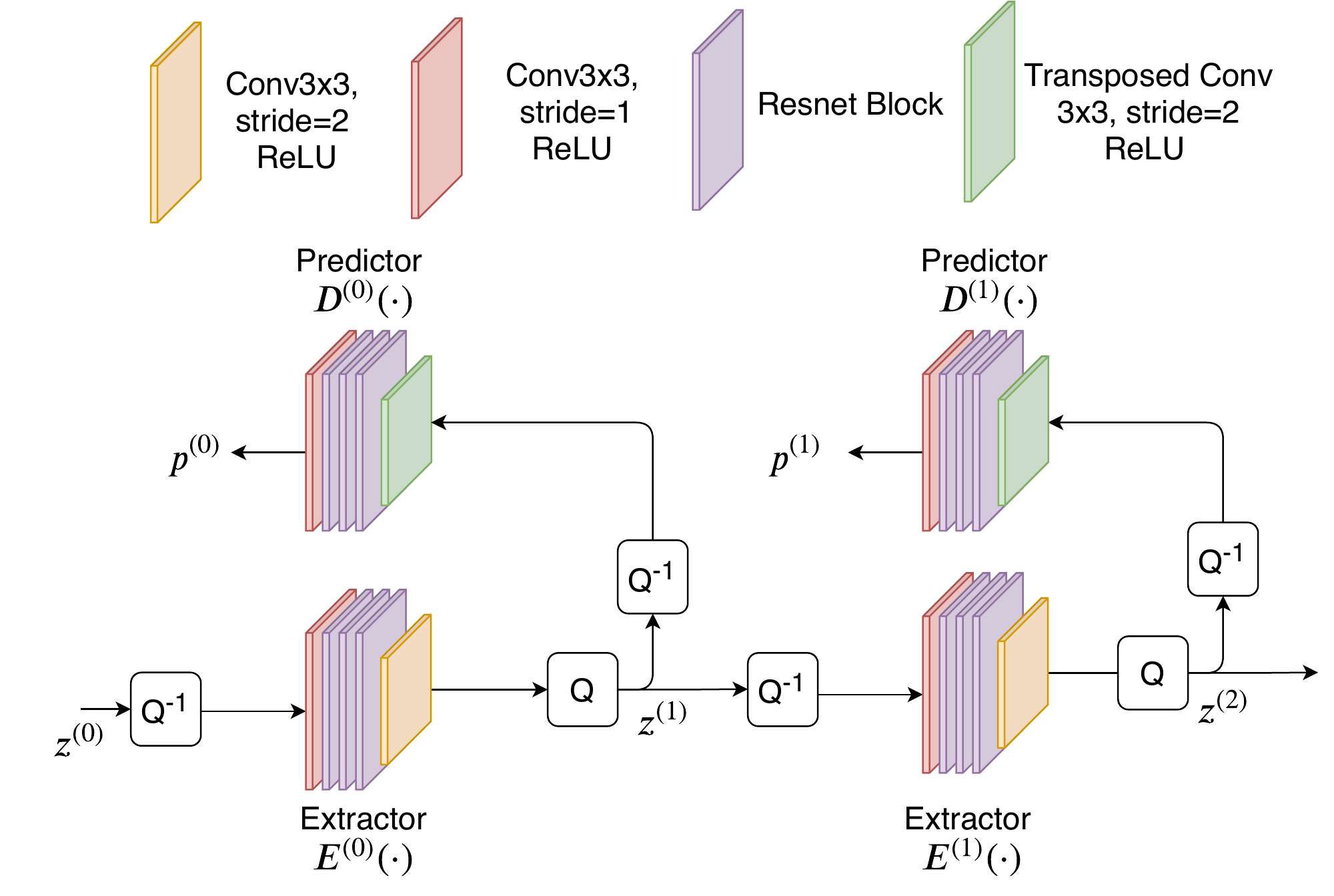}
\caption{Multi-scale Probability Model Architecture\label{fig:Multi-scale-Probability-Model}}
\end{figure}

Figure \ref{fig:Multi-scale-Probability-Model} shows the architecture
of the multi-scale probability model (MSProb). $z^{(0)}$ is the input
of the MSProb model. At scale $i$, MSProb model uses extractor $E^{(i)}$
to generate a downscaled representation $z^{(i+1)}$. Then, predictor
$D^{(i)}$ is applied to calculate the parameters $p^{(i)}$, which
contains parameters $\tilde{\mu}^{(i)}$, $\lambda^{(i)}$, $s^{(i)}$,
and $\pi^{(i)}$ of the discretized logistic mixture model. The same
procedure is applied to every scale. Let $S$ be the number of scales
of the MSProb model. The output of the last scale $z^{(S)}$ is either
uncompressed or compressed using the method provided by the NumPy
package in Python software depending on whichever comes with a smaller
size. At the training stage, outputs $p^{(0)},p^{(1)},\cdots p^{(S-1)}$
are used to calculate the cross-entropy of the output tensors $z^{(0)},z^{(1)},\cdots,z^{(S-1)}$
and the sum of these cross-entropies is taken as the compression loss $\mathcal{R}(z)$.
At the encoding stage, $p^{(i)}$ is used as the probabilities to
encode tensor $z^{(i)}$ using the Arithmetic Encoder. The encoded
bitstreams are put into the output bitstream in the order of $z^{(S)},z^{(S-1)},\cdots,z^{(0)}$.
At the decoding stage, $z^{(S)}$ is first decoded from the bitstream.
Then predictor $D^{(S-1)}(\cdot)$ is used to calculate $p^{(S-1)}$
from $z^{(S)}$ and $z^{(S-1)}$ is decoded from the bitstream. This
procedure is repeated until we have $z^{(0)}$ decoded. We only need
to run the predictor networks $S-1$ times to decode the whole stream.
Our experiments show that a large number of scales does not improve
the performance much. To have a small model, we only use two scales
in our system. Our extractor network halves the width, height and
the number of channels of the input tensors at each scale. We set
the number of mixtures in the mixture model to 5. 


\subsection{Training Objective}
We train our model by using the following objective:
\begin{multline}
\label{eq:loss}
  \mathcal{L}(\hat{f}_t, f_t, z, \tau) = \lambda_1 \mathcal{D}_1(\hat{f}_t, f_t) \\ + \lambda_2 \mathcal{D}_2(\hat{f}_t, f_t) + \lambda_3 \mathcal{R}(z) + \lambda_4 \mathcal{M}(\tau),
\end{multline}
where $\mathcal{D}_1$ is the negative multi-scale structural similarity (MS-SSIM) \cite{Wang2003a}, $\mathcal{D}_2$ is the mean-squared error (MSE), $\mathcal{R}$ is the rate-loss provided by the probability model, $\mathcal{M}$ is the constraint on the importance map defined in Eq. \eqref{eq:im_loss}. $\lambda_1$, $\lambda_2$, $\lambda_3$, $\lambda_4$ are scalar values that are determined empirically.

\section{Experiments}
\label{sec:experiments}
In this section we describe the experimental setup and results.
The number of channels in the convolutional layers for the \textit{frame-encoders} $\Psi$ is $20, 40, 40$, for the \textit{embedding-encoder} $\Phi_E$ is $80, 40, 10$, for the importance map $\Upsilon$ is $40, 20, 1$, for the \textit{embedding-decoder} $\Phi_D$ is $40, 80, 40$, for the \textit{frame-decoder} $\Theta$ is $20, 3, 3$, for the attention network $\Lambda$ is $40, 3, 3$, and for the final layer $\rho$ is $3$. 
The training was performed on full-resolution frames from the CLIC training dataset for the P-frame compression track, using a batch-size of $144$, and the Adam optimizer. During training, we gradually decreased the learning rate from $0.001$ to $0.0002$, increased $\lambda_3$ from $0.0001$ to $0.001$, increased $\lambda_4$ from $0.0001$ to $0.5$, and decreased $\beta$ from $0.5$ to $0.3$. $\lambda_1$ and $\lambda_2$ were set to $1.0$. Training was performed for $13$ epochs. 

When evaluated on the CLIC validation dataset, our model achieves MS-SSIM of $0.978$, Peak Signal-to-Noise Ratio (PSNR) of $30.44 dB$, bits-per-pixel (BPP) of $0.0707$. 
The decoder size is about $15.8$MB and decoding time is 1484 seconds, making our system one of the most memory and computationally efficient among all entries to the CLIC challenge.

We also performed an ablation study where we excluded in turn the importance maps and the attention mechanism. This ablation study was done by training and testing on 10 videos out of the total 733 videos, selected so as to belong each to a different content type (e.g., \textit{Animation}, \textit{Gaming}, \textit{VR}, \textit{Lecture}, \textit{MusicVideo}, \textit{Sport}). The hyper-parameters are: batch-size $32$, learning rate $0.001$, $\lambda_1, \lambda_2=1.0$, $\lambda_3=0.0001$, $\lambda_4=0.01$, $\beta=0.3$. The comparison with respect to the full model was done at $40K$ training iterations and is reported in Table \ref{table1}.  As can be seen from the table, attention is necessary to achieve higher MS-SSIM and PSNR, while it also helps in decreasing the BPP. Removing the importance maps significantly deteriorates the compression rate.

\begin{table}[h!]
  \begin{center}
    \caption{Ablation study on a subset of CLIC dataset.}
    \label{table1}
    \begin{tabular}{|c|c|c|c|} 
    \hline
      \textbf{Model} & \textbf{MS-SSIM} & \textbf{PSNR} & \textbf{BPP} \\
      \hline
      \hline
      Full & $0.955$ & $30.35$ & $6.6\mathrm{e}{-3}$ \\
      \hline
      No attention & $0.949$ & $28.70$ & $8.7\mathrm{e}{-3}$ \\
      \hline
      No importance maps & $0.960$ & $31.23$ & $13.8\mathrm{e}{-3}$ \\
      \hline
      \vtop{\hbox{\strut No importance maps}\hbox{\strut @MS-SSIM=$0.955$}} & $0.955$ & $30.57$ & $13.4\mathrm{e}{-3}$ \\
      \hline

    \end{tabular}
  \end{center}
\end{table}

\section{Conclusions}
\label{sec:conclusions}
In this paper, we proposed an end-to-end learned model for compressing video frames. We compute differences between frames in embedding-space, which are then analyzed in order to compute the importance of different parts of the tensor to be encoded. At decoder-side we designed an attention mechanism to adaptively combine an initial predicted current frame and the previous frame. In our experimental section, we showed the effectiveness of our approach and we highlighted the contribution of different components. 
The authors would like to thank the following colleagues for their valuable help and discussions: Yat Lam, Alireza Zare, Goutham Rangu, Yu You.

{\small
\bibliographystyle{ieee}
\bibliography{egbib,context_model_clean}

\begin{thebibliography}{10}\itemsep=-1pt

\bibitem{balle2018variational}
J.~Ball{\'e}, D.~Minnen, S.~Singh, S.~J. Hwang, and N.~Johnston.
\newblock Variational image compression with a scale hyperprior.
\newblock {\em arXiv:1802.01436 [cs, eess, math]}, May 2018.
\newblock arXiv: 1802.01436.

\bibitem{Chen2019a}
T.~Chen, H.~Liu, Z.~Ma, Q.~Shen, X.~Cao, and Y.~Wang.
\newblock Neural image compression via non-local attention optimization and
  improved context modeling.
\newblock arXiv:1910.06244, 2019.

\bibitem{chen2019learning}
Z.~Chen, T.~He, X.~Jin, and F.~Wu.
\newblock Learning for {Video} {Compression}.
\newblock {\em IEEE Transactions on Circuits and Systems for Video Technology},
  pages 1--1, 2019.
\newblock arXiv: 1804.09869.

\bibitem{choi2019deepframe}
H.~Choi and I.~V. Bajic.
\newblock Deep {Frame} {Prediction} for {Video} {Coding}.
\newblock {\em arXiv:1901.00062 [cs, eess]}, June 2019.
\newblock arXiv: 1901.00062.

\bibitem{Cricri2016}
F.~Cricri, X.~Ni, M.~Honkala, E.~Aksu, and M.~Gabbouj.
\newblock Video ladder networks.
\newblock {\em ArXiv}, abs/1612.01756, 2016.

\bibitem{LiM2018}
M.~Li, W.~Zuo, S.~Gu, D.~Zhao, and D.~Zhang.
\newblock Learning convolutional networks for content-weighted image
  compression.
\newblock In {\em Proceedings of the IEEE Conference on Computer Vision and
  Pattern Recognition}, pages 3214--3223, 2018.

\bibitem{Mentzer2018}
F.~Mentzer, E.~Agustsson, M.~Tschannen, R.~Timofte, and L.~V. Gool.
\newblock Conditional probability models for deep image compression.
\newblock {\em 2018 IEEE/CVF Conference on Computer Vision and Pattern
  Recognition}, pages 4394--4402, 2018.

\bibitem{mentzer2019practical}
F.~Mentzer, E.~Agustsson, M.~Tschannen, R.~Timofte, and L.~Van~Gool.
\newblock Practical {Full} {Resolution} {Learned} {Lossless} {Image}
  {Compression}.
\newblock {\em arXiv:1811.12817 [cs, eess]}, May 2019.
\newblock arXiv: 1811.12817.

\bibitem{oord2016pixelrecurrent}
A.~v.~d. Oord, N.~Kalchbrenner, and K.~Kavukcuoglu.
\newblock Pixel {Recurrent} {Neural} {Networks}.
\newblock {\em arXiv:1601.06759 [cs]}, Aug. 2016.
\newblock arXiv: 1601.06759.

\bibitem{Salimans2017}
T.~Salimans, A.~Karpathy, X.~Chen, and D.~P. Kingma.
\newblock Pixelcnn++: Improving the pixelcnn with discretized logistic mixture
  likelihood and other modifications.
\newblock {\em ArXiv}, abs/1701.05517, 2017.

\bibitem{Shi2016}
W.~Shi, J.~Caballero, F.~Husz{\'a}r, J.~Totz, A.~P. Aitken, R.~Bishop,
  D.~Rueckert, and Z.~Wang.
\newblock Real-time single image and video super-resolution using an efficient
  sub-pixel convolutional neural network.
\newblock {\em 2016 IEEE Conference on Computer Vision and Pattern Recognition
  (CVPR)}, pages 1874--1883, 2016.

\bibitem{Sullivan2012a}
G.~Sullivan, J.-R. Ohm, and T.~Wiegand.
\newblock Overview of the high efficiency video coding (hevc) standard.
\newblock {\em Circuits and Systems for Video Technology, IEEE Transactions
  on}, 22, 12 2012.

\bibitem{Theis2017}
L.~Theis, W.~Shi, A.~Cunningham, and F.~HuszÃ¡r.
\newblock Lossy image compression with compressive autoencoders.
\newblock In {\em International Conference on Learning Representations}, 03
  2017.

\bibitem{Wang2003a}
Z.~{Wang}, E.~P. {Simoncelli}, and A.~C. {Bovik}.
\newblock Multiscale structural similarity for image quality assessment.
\newblock In {\em The Thirty-Seventh Asilomar Conference on Signals, Systems
  Computers, 2003}, volume~2, pages 1398--1402 Vol.2, 2003.

\end{thebibliography}
}

\end{document}